\documentstyle[psfig,12pt,qqa4lart]{article}
\input tcilatex

\QQQ{Language}{
American English
}

\begin{document}

\author{V. R. Chitnis and P. N. Bhat \\ 
{\it Tata Institute of Fundamental Research, }\\{\it   
 Homi Bhabha Road,
 Mumbai 400 005, India.}}

\title{\v Cerenkov Photon Density Fluctuations in Extensive Air Showers}
\maketitle
\begin{abstract}
The details of \v Cerenkov light produced by a $\gamma -$ ray or a cosmic
ray incident at the top of the atmosphere is best studied through systematic 
simulations of the extensive air showers. Recently such
studies have become all the more important in view of the various techniques
resulting from such studies, to distinguish $\gamma -$ ray initiated
showers from those generated by much more abundant hadronic component of
cosmic rays. We have carried out here such systematic simulation studies
using CORSIKA package in order to understand the \v Cerenkov photon density
fluctuations for 5 different energies at various core distances both for
$\gamma -$ ray and proton primaries incident vertically at the top of the
atmosphere. Such a systematic comparison of shower to shower density
fluctuations for $\gamma -$ ray and proton primaries is carried out for the
first time here. It is found that the density fluctuations are significantly
non-Poissonian. Such fluctuations are much more pronounced in the
proton primaries than $\gamma -$ ray primaries at all energies. 
The processes that contribute significantly to the observed fluctuations 
have been identified. It has been found that significant contribution to
fluctuations comes from photons emitted after shower maximum. 
The electron number fluctuations and correlated emission of \v Cerenkov
photons are mainly responsible for the observed fluctuations.
\end{abstract}

\section{Introduction}

Ground based atmospheric \v Cerenkov technique is, at present, 
the only way by which TeV $\gamma -$ rays could be detected from point 
sources such as $\gamma -$ ray
pulsars, short period X$-$ ray binaries or BL-Lac objects. Recent detection
of TeV emission from a few of these objects (Vacanti {\it et al.,} 1991; Punch 
{\it et al.,} 1992, Chadwick {\it et al., }1997, Quinn {\it et al.,} 1997, 
Weekes, 1988, Fegan, 1994) has
created much interest in the field of TeV $\gamma -$ ray astronomy. The $%
\gamma -$ ray signals found typically are $\sim 1\%$ of the more abundant
background events of cosmic ray nuclei, particularly protons. In order to
detect faint Very High Energy (VHE) $\gamma -$ ray sources, one has to
improve the signal to noise ratio by rejecting the bulk of the hadronic
background. In order to do so it is imperative to study the detailed
characteristics of \v Cerenkov light production by photon initiated and
proton initiated cascades in the atmosphere.

Simulation studies in the past (Rao \& Sinha, 1988; Hillas \& Patterson,
1987; Zatsepin \& Chudakov, 1962) have shown that the \v Cerenkov pool at
the observation level has the signature of the primary. The lateral
distribution of \v Cerenkov radiation seems to be distinctly different in $%
\gamma -$ ray and proton initiated showers in the sense that in the former
case it is flat upto about 140 m and characterized by an increased photon
density (called the `hump') at that distance while in the latter case it is
steeper with practically no hump. It has been suggested (Rao \& Sinha, 1988)
that this characteristic difference could be measured in an observation and
could be used for improving the signal to noise ratio. These arguments are
based on the average properties of showers. In practice, however, \v
Cerenkov photon density fluctuations play a dominant role. 
The signature of the primary gets buried in the noise which is mainly due to
large fluctuations in photon densities from shower to shower, even at same
energy. These fluctuations in turn reduce the efficiency by which the
primary could be identified based on the lateral distribution measurements
(Krys and Wasilewski, 1993). As a result, the study of the photon
fluctuations plays an important role in deciding the signal to noise ratio
of the VHE $\gamma -$ ray observations based on the
measurement of lateral distribution of \v Cerenokov light.

It has been known that shower to shower fluctuations in proton initiated
cascades are expected to be much larger than $\gamma -$ ray initiated
cascades since the nuclear interaction mean free path (70 $g$ $cm^{-2}$) is
about twice as large as the radiation length (37.15 $g$ $cm^{-2}$) 
in the atmosphere and the
number of secondaries and their energy spectra are known to fluctuate
widely. Furthermore muons, which are present only in hadron initiated
showers (above the \v Cerenkov threshold $E_\mu \sim 4$ GeV) reach the
observation level and could create local peaks in the light pool at the
observation level (Hillas \& Patterson, 1987). In this paper we make an
attempt to estimate and compare the extent of these fluctuations in proton
and $\gamma -$ ray initiated showers at various energies and estimate the 
relative contributions from different but known sources of fluctuations.

\section{The Simulations}

We have used CORSIKA package version 4.502 (Knapp \& Heck, 1995) for
simulation of air showers generated by $\gamma -$ rays and protons.
This package simulates interactions of nuclei, hadrons, muons, electrons 
and photons as well as decays of unstable secondaries in the atmosphere.
It also provides information about the type, energy, location, direction 
and arrival times of all the secondary particles generated in the air shower, 
which reach the observation level. It also supports the
option of generating \v Cerenkov photons emitted by various particles in the
shower. It uses the EGS4 package (Nelson {\it et al., }1985) for the
development of the electromagnetic cascade in the atmosphere. The \v
Cerenkov radiation produced by the secondary charged particles within the
specified bandwidth (300-550 nm) is propagated to the ground. The
position, angle, time and production height of each \v Cerenkov photon
hitting the detector at the observation level are recorded. However, the
wavelength dependent absorption of these photons in the atmosphere is not
taken into account.

In the present work we have studied the \v Cerenkov light produced by
monoenergetic $\gamma -$ rays and protons incident vertically at the top of
the atmosphere. Height of the first interaction is selected randomly taking
into consideration the appropriate mean free path. Target of the first
interaction is chosen randomly according to atmospheric abundances. In the
simulations, we have considered an array of detectors with area 2.11 $\times $
2.11 m$^2$ (this corresponds to the total mirror area of each telescope in
the Pachmarhi Array of \v Cerenkov Telescopes (PACT), see Bhat et al. 1997), 
with
17 detectors in X-direction and 21 detectors in Y-direction. Spacing between
the detectors in X-direction is 25 m and in Y-direction is 20 m. It may be
mentioned that even though PACT consists of only 25 telescopes each of area
4.45 m$^2$, a larger array is chosen for simulation purposes in order to
study core distance dependent properties of PACT. Only those \v
Cerenkov photons which hit any of the detectors of the array at all 
incident angles are recorded.
Observation altitude and magnetic field appropriate for Pachmarhi
(longitude: 78$^{\circ }$ 26$^{\prime }E$; latitude: 22$^{\circ }$ 28$%
^{\prime }N$ and altitude: 1075 $m$ amsl) location are taken into account. 
The shower core is always chosen to be at the centre of the array.

An option of variable bunch size is available in the package, which defines
the number of \v Cerenkov photons treated together. This will also reduce
the size of the output data file. However, it was noticed that for
fluctuation studies the bunch size has to be set to unity to get the correct
estimate of the fluctuations as larger bunch size tends to overestimate the
photon fluctuations.

\section{Average \v Cerenkov lateral distributions}

\v Cerenkov photon lateral distributions were generated and averaged over
several showers. For various energies of $\gamma -$ ray and proton
primaries, typically 100 showers were generated. For a few energies larger
number of showers were generated (e.g. 200 showers for 100 GeV 
$\gamma -$ rays and 400 showers for 50 GeV $\gamma -$ rays and 150 GeV 
protons) to ensure that average shape of the lateral distribution does not
critically depend on the sample size. For 1 TeV $\gamma -$ rays and 2 TeV 
protons 50 showers were generated. Fig. 1a shows the
average lateral distributions of \v Cerenkov photons generated by $\gamma -$
rays of energies 50, 100, 250, 500 and 1000 GeV. Corresponding distribution
for protons of energies 150 GeV, 250 GeV, 500 GeV, 1000 GeV and 2000 GeV are
shown in fig. 1b. \v Cerenkov photon densities shown in the plots are
averaged over 10 consecutive detectors when arranged with increasing core
distance. Proton energies are chosen such that their \v Cerenkov yield is
comparable to that of $\gamma -$ rays. Even though the average \v Cerenkov 
lateral distribution shown in the fig. 1 is derived for the above mentioned 
detector array, it is verified to be independent of detector size and spacing.

It may be seen that lateral distributions produced by $\gamma -$ ray
primaries show a characteristic hump at a distance of about 135 m from the
core for energies of atleast up to 1 TeV. The origin of a hump in the case of $%
\gamma -$ ray primaries is discussed in detail by Rao \& Sinha (1988). This
is due to the focussing of \v Cerenkov photons from a large range of
heights, over which the product of height and \v Cerenkov angle (h$\theta _c$%
) is approximately constant. It has also been demonstrated by Rao \& Sinha 
that only the
higher energy ($\geq $ 1 GeV) electrons are responsible for the production
of the hump. The cumulative RMS scattering angle of these electrons, 
(which is inversely proportional to the electron energy), is smaller than 
the \v Cerenkov angle.
Whereas in the case of lower energy electrons this angle is larger than the
\v Cerenkov angle. This enables \v Cerenkov photons to reach the regions
on either side of the hump. In case of proton primaries,
electrons are pair produced by $\gamma -$ rays, which are decay products
of $\pi ^0$s, and have their production angles determined by the transverse
momentum of $\pi ^0$s. This production angle is also to be taken into
consideration while calculating the threshold energy of electrons, at 
which \v Cerenkov angle is equal to the scattering angle. Hence, this
threshold energy for proton primaries is larger than the corresponding
value for $\gamma -$ ray primaries, consequently reducing the relative 
number of electrons above this threshold. As a result these electrons do not
produce a noticeable hump. 

Prominence of the hump reduces as the energy of $\gamma -$ ray primary 
increases
from 50 GeV to 1 TeV, since higher energy electrons penetrate deeper in the
atmosphere, thereby increasing the contribution to \v Cerenkov radiation
from electrons from lower altitudes where h$\theta _c$ starts decreasing
(Rao \& Sinha, 1988). 
\begin{figure}
\centerline{\psfig{file=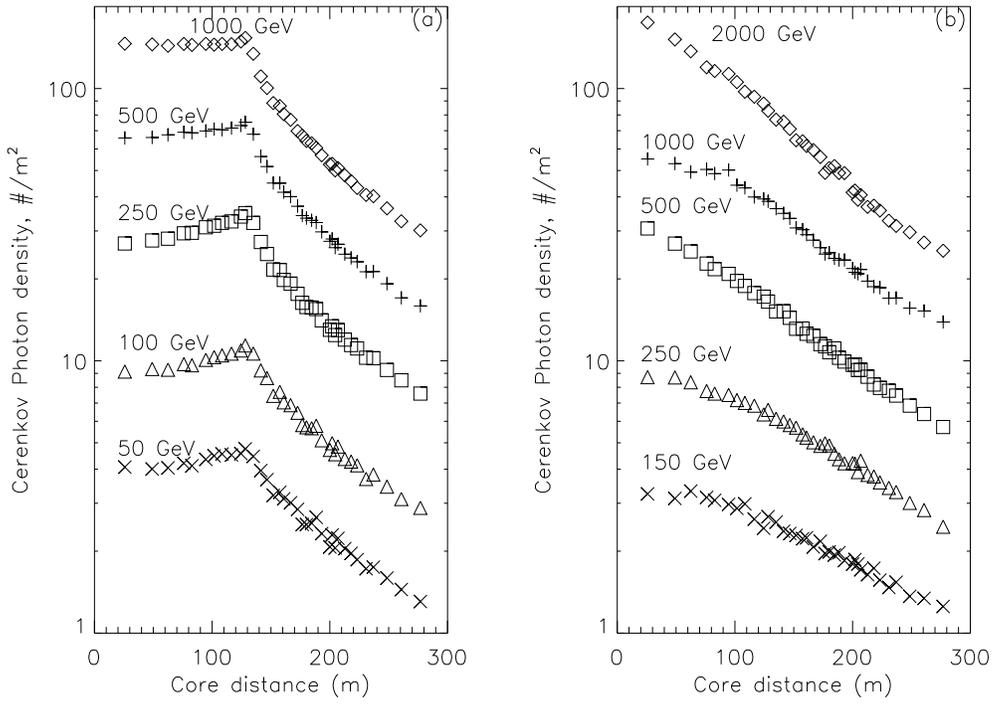,height=10cm}}
\caption{Average lateral distributions of \v Cerenkov photon densities 
resulting from extensive air showers initiated by vertically incident 
(a)$\gamma -$ rays and (b)protons of various energies. Average of 400 
showers is used for 50 GeV $\gamma -$ rays and 150 GeV protons, 200 
showers for 100 GeV $\gamma -$ rays, 100 showers for the rest except for
1 TeV $\gamma -$ rays and 2 TeV protons for which 50 showers each were simulated.}  
\end{figure}
In order to study the variation of the strength of the hump as a function of
primary energy we define the strength of the hump as the ratio of the
density at the hump to that at the shower core. We generated nearly 300 $%
\gamma -$ ray showers of varying energies in the range 50-1000 GeV and computed
this ratio for each shower. Fig 2 shows the variation of this ratio with
primary $\gamma -$ ray energy. The ratio decreases with increasing energy as
expected and a power law with a slope of -0.38 fits the data well showing
that eventually this ratio reduces to a limiting value of 1.

\begin{figure}
\centerline{\psfig{file=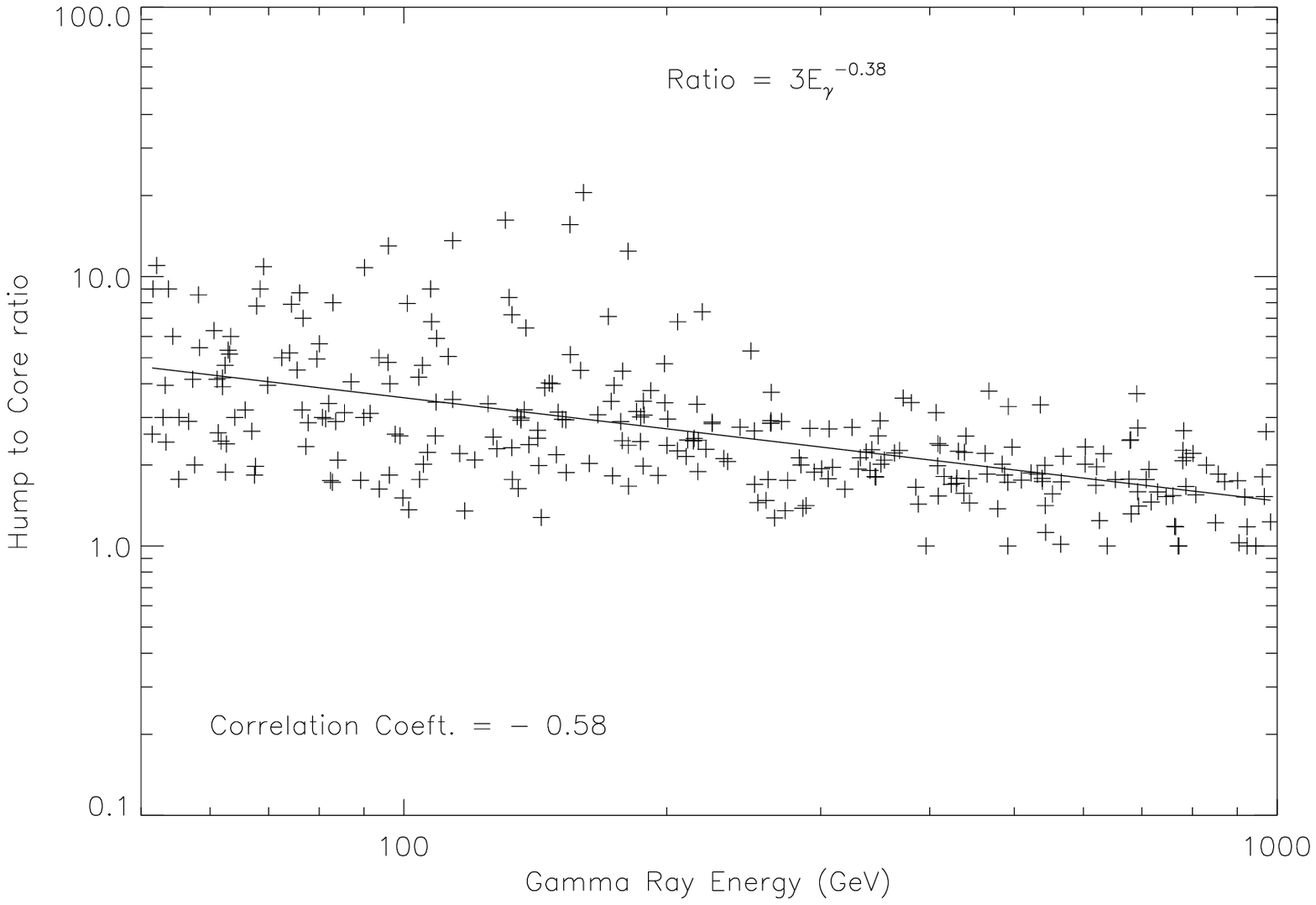,height=7cm}}
\caption{ Variation of the ratio of the photon density at the hump to that 
at the core with primary $\gamma -$ ray energy. There seems to be a good 
anticorrelation with 
energy while the ratio when fitted with a power law has a slope -0.38}
\end{figure}

\section{Shower size fluctuations}

\begin{figure}
\centerline{\psfig{file=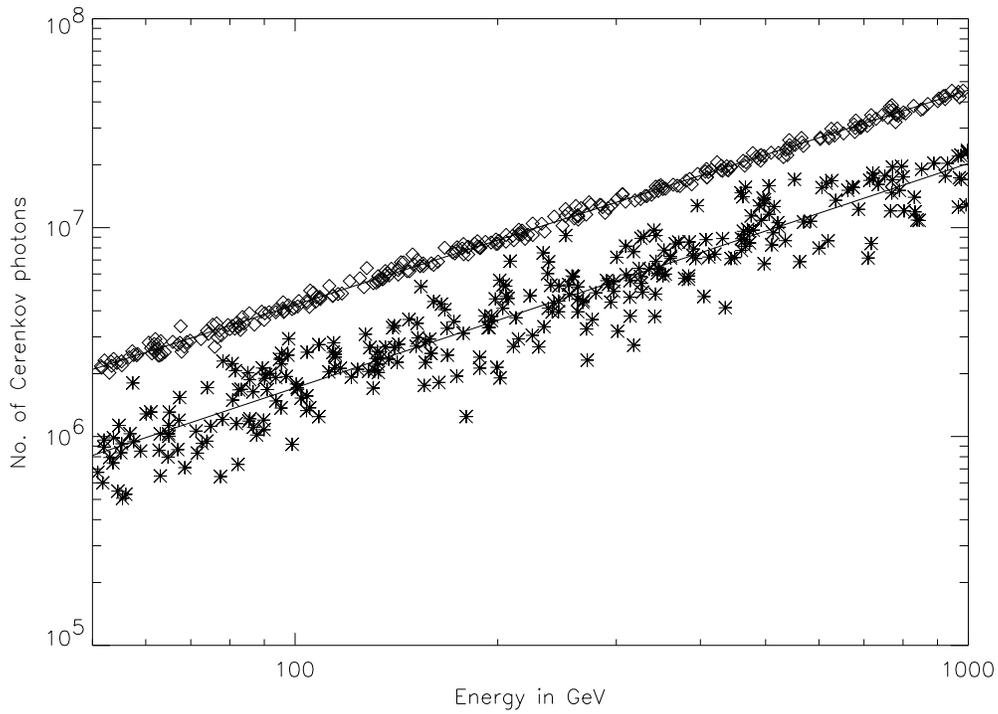,height=10cm}}
\caption{ Variation of the total number of \v Cerenkov photons in the 
bandwidth 300-550 nm produced by vertically incident 
$\gamma -$ rays (diamonds) and protons (stars) with primary energies, 
indicating the 
magnitude of photon number fluctuations. The proton generated showers exhibit 
a significantly higher degree of fluctuations as compared to those by 
$\gamma -$ rays.  } 
\label{Figure 3:}
\end{figure}

Figure 3 shows the total number of \v Cerenkov photons produced in the
atmosphere by showers initiated by vertically incident $\gamma -$ rays and 
protons of various energies in the range 50 - 1000 GeV, with an assumed energy
spectrum of the form, $E^{-1}$. Nearly 300 showers each were generated for 
$\gamma -$ ray and proton primaries. While there is a very good
proportionality between the shower size (hereafter measured in terms of the
total number of \v Cerenkov photons produced in a shower) and primary energy
in the case of $\gamma -$ ray primaries (linear correlation coefficient = 
0.998), it is not as good in the case of proton primaries (linear correlation 
coefficient = 0.9), due to increased fluctuations in the latter case. The best 
power-law fit parameters to the total number of \v Cerenkov photons 
as a function of the primary energy are shown in table 1.

\begin{table} 
\caption{Least squares fit coefficients for a linear fit to the log of total number of
\v Cerenkov photons as a function of the log of the primary energy for  vertically
incident $\gamma -$ rays and protons.}
  \begin{tabular}{lll} \hline
     {\bf Primary}  &  {\bf Intercept}  &  {\bf Slope} \\
     {\bf Species}  &   &   \\ \hline
    $\gamma -$ rays       &  10.53 $\pm $ 0.02    &  1.028 $\pm $ 0.003   \\  
     Protons      &  9.41   $\pm $ 0.11   &  1.07 $\pm $ 0.02  \\ \hline
\end{tabular}
\end{table}

\begin{figure}
\centerline{\psfig{file=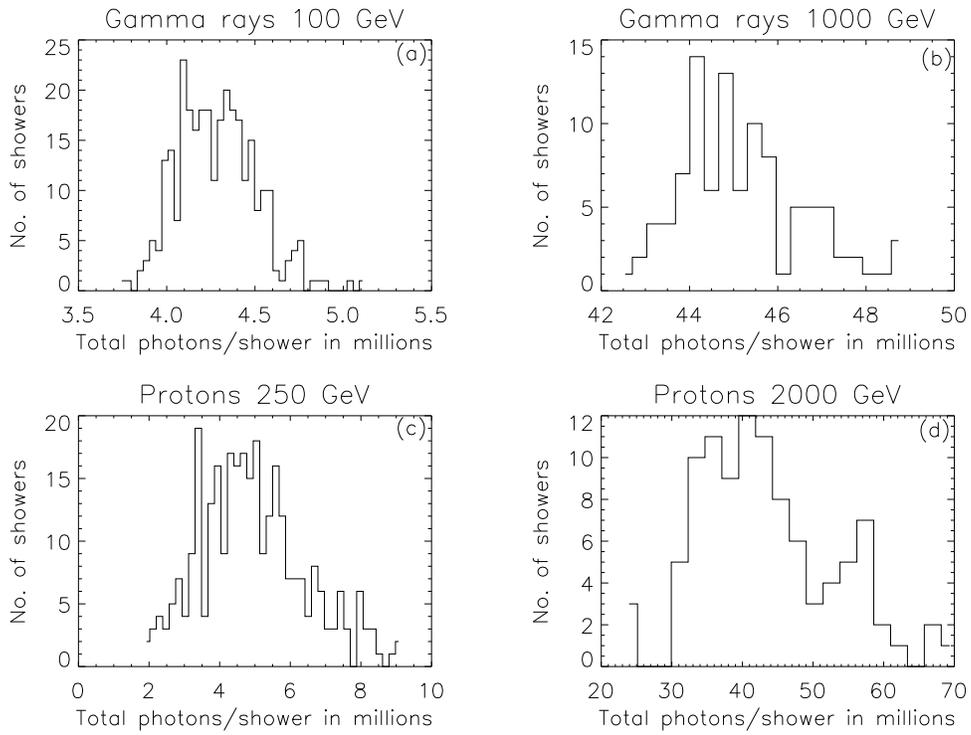,height=10cm}}
\caption{ The distributions of the total number of \v Cerenkov photons in the 
bandwidth 300-550 nm in showers generated by vertically incident $\gamma -$ 
rays and protons each with two different primary energies.} 
\label{Figure 4:}
\end{figure}

Distributions of the total number of \v Cerenkov photons produced by $\gamma
-$ ray primaries of energy 100 GeV and 1 TeV and proton primaries of
energies 250 GeV and 2 TeV are shown in fig. 4 a-d respectively. Because of
the higher \v Cerenkov yield from $\gamma -$ ray primaries (Browning and Turver,
1977), higher energy protons with comparable \v Cerenkov yields have been
chosen for comparison. We have generated 300 showers for lower energy and
100 showers for higher energy $\gamma -$ rays and protons. \v Cerenkov
photon distributions are much broader for proton primaries compared to those
from $\gamma -$ rays, consistent with fig. 3. The ratio of rms deviation to 
average number of \v
Cerenkov photons decreases with increasing energy for both $\gamma -$ rays
and protons. When the $\gamma -$ ray energy increases from 100
GeV to 1000 GeV this ratio falls from 5\% to 3\%. Similarly when the
proton energy increases from 250 to 2000 GeV this ratio falls from 30\%
to 23\%. These results are consistent with those obtained
by Ong et al. (1995) (see section 8.2 for details). 

\section{Density fluctuations}

\subsection{Intra-shower fluctuations}

\begin{figure}
\centerline{\psfig{file=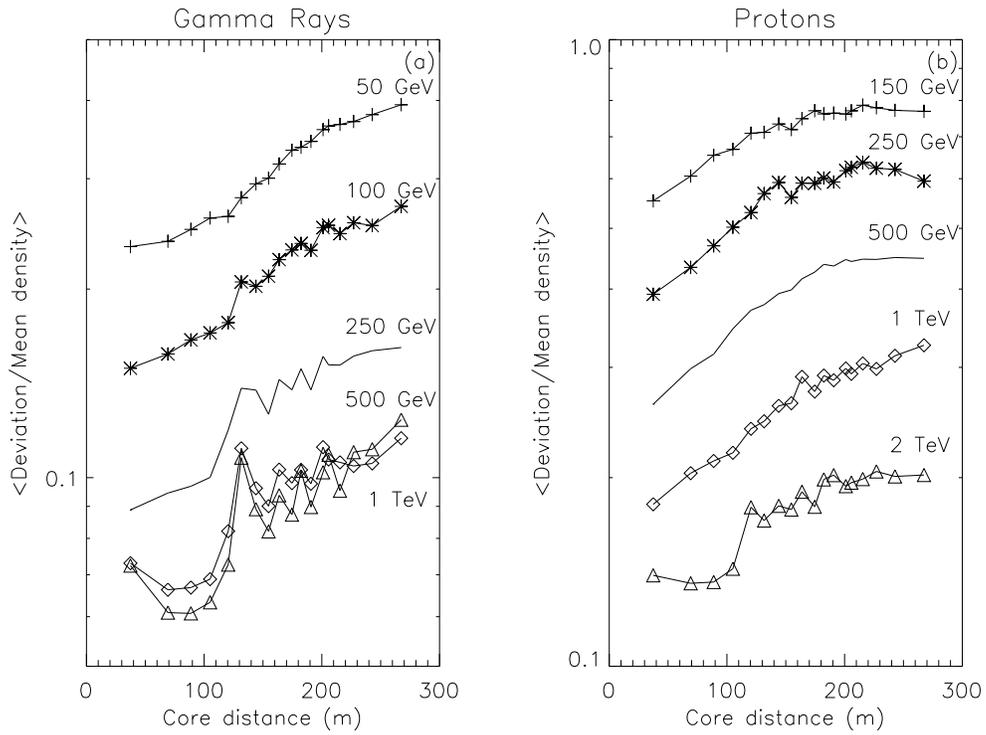,height=10cm}}
\caption{Ratio of the magnitude of deviation to the mean photon density 
as a function of core distance for (a) $\gamma -$ rays and (b) protons. 
Mean densities are calculated for groups of 10
consecutive detectors arranged with increasing core distance. Ratios
are averaged over 400 showers for 50 GeV $\gamma -$ rays and 150 GeV
protons, 200 showers for 100 GeV $\gamma -$ rays, 50 showers for 1 TeV
$\gamma -$ rays and 2 TeV protons. 100 showers are used for the rest.}
\label{Figure 5:}
\end{figure}

Figure 5 shows the variation of the ratio of the absolute deviation to the mean photon 
density as a function of core distance, averaged over a number of
showers. For each shower, average \v Cerenkov photon densities are
calculated over groups of 10 consecutive detectors, when arranged with 
increasing core distance. Absolute value of the difference
between individual photon densities and corresponding mean 
density is defined as the deviation. Ratio of deviation to the mean density 
is calculated for
each detector, for each shower. Ratios are then averaged over total number
of showers (fig. 5).
In the case of both $\gamma -$ rays and protons, ratio of the deviation
to the mean photon density increases with core distance. For a given primary,
it decreases with increase in the energy of the primary. It is higher
by a factor of about 2 for protons of energies in the range 150 GeV - 2 TeV, 
compared to those for $\gamma -$ rays of 
comparable \v Cerenkov yields. At higher $\gamma -$ ray energies, i.e.,
500 GeV and 1 TeV, there is a sharp increase in the ratio in the hump region.
Similar behaviour is also seen in the case of 2 TeV protons. 

This ratio can be measured in an atmospheric \v Cerenkov experiment. In
principle, this  
could be used to identify the showers generated by $\gamma -$ ray primaries
if the energy is independently estimated. 

In rest of the paper we will be discussing the inter-shower photon density
fluctuations only.

\begin{figure}
\centerline{\psfig{file=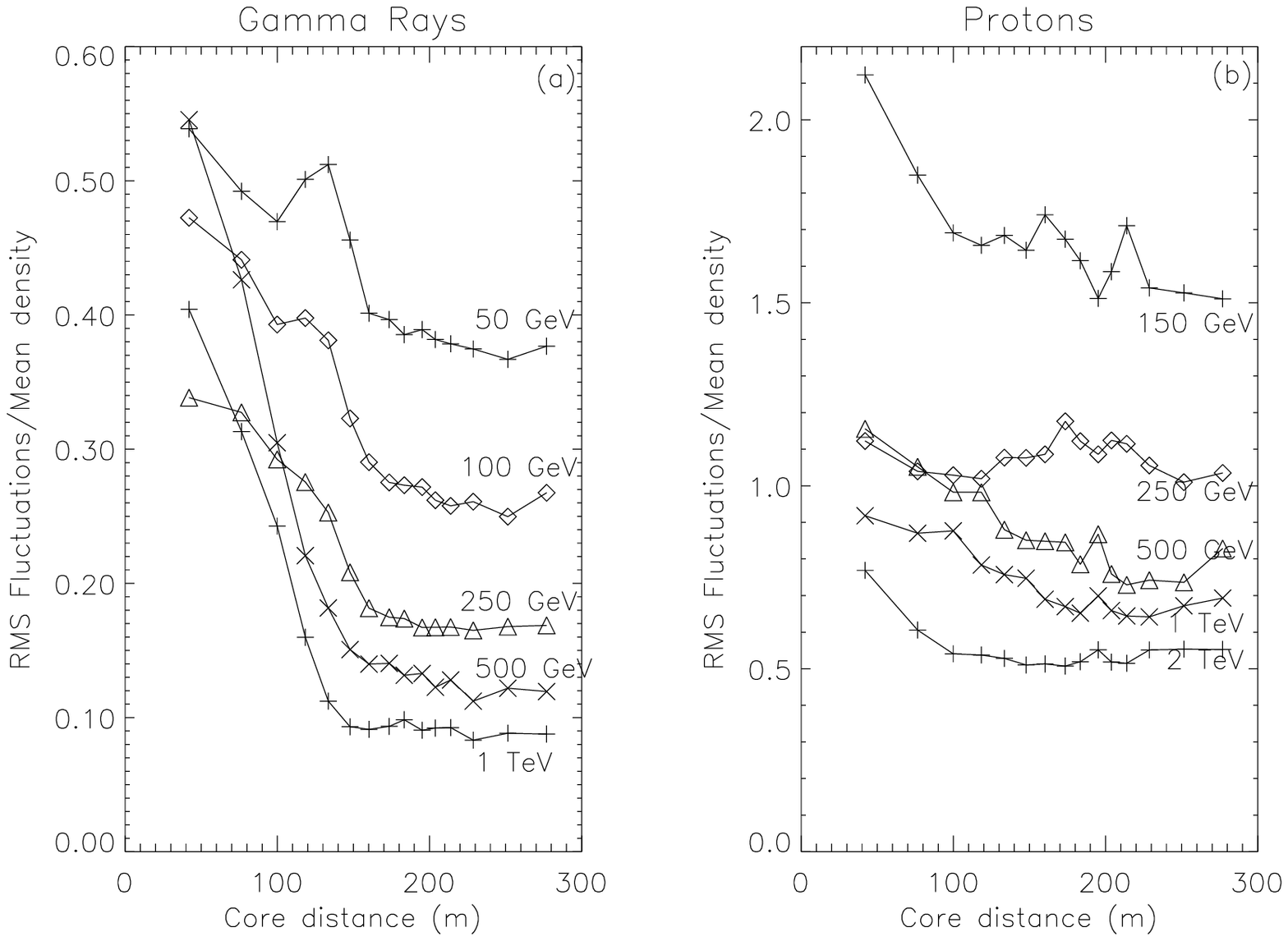,height=10cm}}
\caption{Ratios of non-statistical RMS values to the mean number of
\v Cerenkov photon densities as a function of core distance at five different
energies of primary (a) $\gamma -$ rays and (b) protons.
The relative fluctuations for proton primaries are higher and  seem to decrease
with increase in  primary energies in cases of both protons and $\gamma -$
rays. }
\label{Figure 6:}
\end{figure}

\begin{figure}
\centerline{\psfig{file=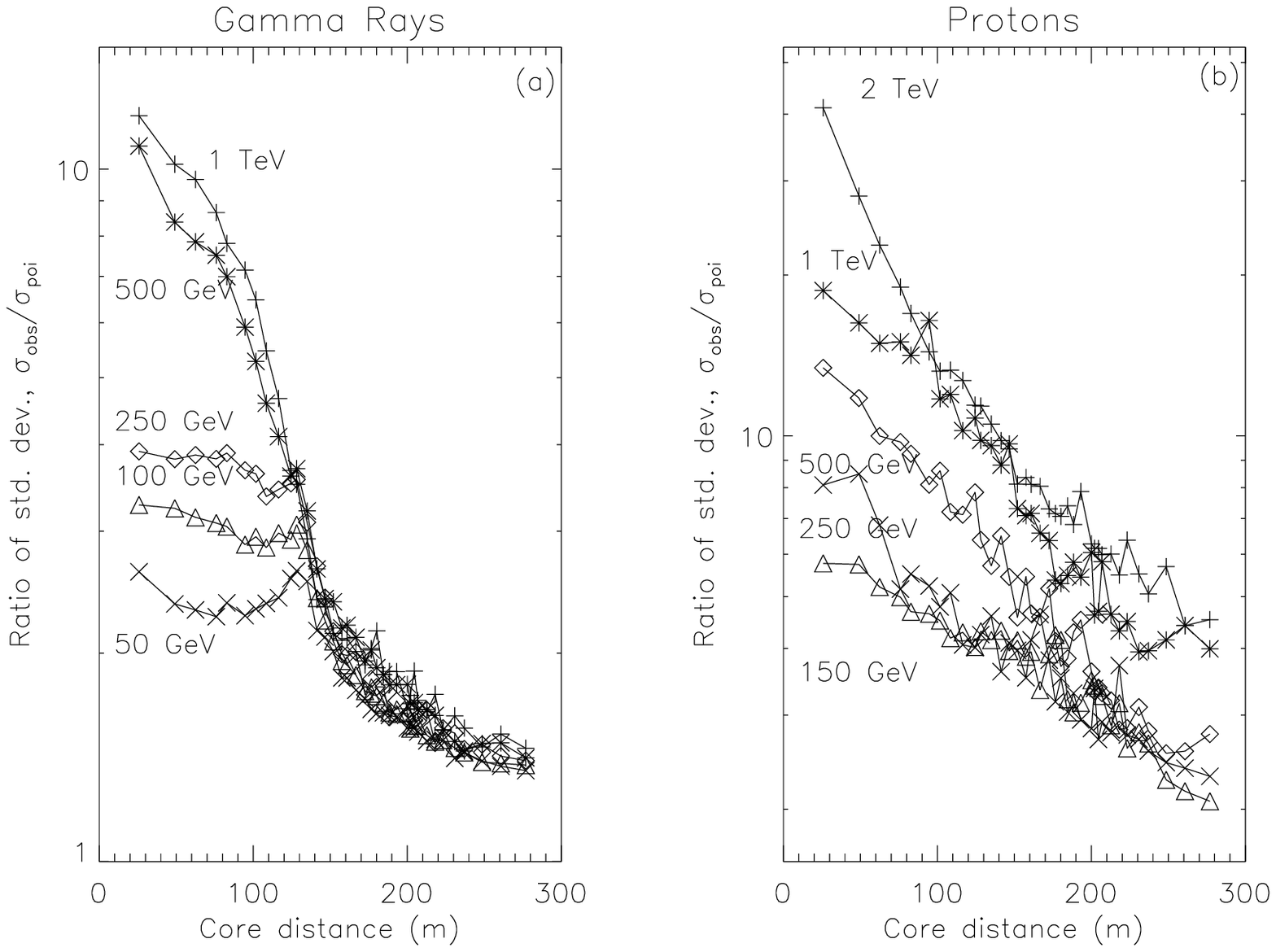,height=7cm}}
\caption{Ratio of the observed RMS fluctuations to Poissonian fluctuations
for (a) $\gamma -$ ray and (b) proton primaries of various energies as a
function of core distance.}
\label{Figure 7:}
\end{figure}

\subsection{Inter-shower fluctuations}

Figure 6 shows the variation of the relative shower to shower photon density 
fluctuations as a function of core distance for 5 different primary 
energies (a) of 50, 100, 250, 500 \& 1000 GeV for $%
\gamma -$ ray primaries and (b) 150, 250, 500, 1000 and 2000 GeV for protons.
The relative fluctuations are measured as a ratio of non-statistical RMS 
values to the mean number of photons detected at a detector of area 4.45 m$^2$.
In general the RMS fluctuations are functions of mean densities. In order
to remove this dependence the non-statistical component of the fluctuations
is estimated assuming that the total fluctuation is given by the quadratic
sum of the statistical (or Poissonian) and non-statistical components.
It has been verified that the ratio of the non-statistical component of
the RMS fluctuations to the mean value of the density over all the showers
($r$) is independent of detector size and spacing. The relative fluctuations at
all primary energies are seen to decrease with increasing core distance 
and reach a constant value beyond the hump region. The increased fluctuations
in the pre-hump region are due to correlated emission from high energy
electrons (see section 7.1 for more details). 

For a given primary the relative fluctuations decrease with increasing 
primary energy as observed by others (Ong, 1995). This is primarily due to 
the reduced electron number fluctuations at higher primary energies. 
The decrease in the degree of fluctuations with increasing primary energy is
monotonic only beyond the hump region. It decreases monotonically up to a 
primary energy of say 250 GeV for $\gamma -$ ray primaries at all core 
distances. It can be seen from fig. 6 that at primary energies of 500 GeV and 
1 TeV the fluctuations within the hump region increase dramatically exceeding
the values at lower energies. Table 2 shows the average production
heights of \v Cerenkov photons at 3 core distance ranges
for 3 different primary energies viz. 100, 500 and 1000 GeV. It can be seen
that the differences in the \v Cerenkov production heights for near core
distances and hump region increase with increasing energy.
This is due to the production of \v Cerenkov photons by the
electrons which survive to this height. At higher energies larger number of
electrons survive closer to the observation level and produce \v Cerenkov
photons directly before undergoing Coulomb scattering and thus contribute
to the increased photon density fluctuations within the hump region. For 
example, the direct 
\v Cerenkov photons reaching the core distance of about 45 m are originated
at a height of 2.2 km ($\sim 800 g cm^{-2}$). At this height, a total of nearly 16 and 44 electrons 
(and positrons) of average energy of 0.04 GeV survive in case of 500 GeV and 
1000 GeV $\gamma -$ ray showers, respectively. They form nearly
3\% and 4\% of the maximum number of electrons produced. Photons 
emitted by some of these electrons
could be correlated and hence contribute significantly to the fluctuations. 

Table 3 shows the average arrival angles of \v Cerenkov photons at different
core distance ranges for 3 different $\gamma -$ ray primary energies. The
increasing average arrival angle with increasing primary energy demonstrates
that \v Cerenkov emission from electrons has a larger lateral 
spread at higher energies.
At core distances past the hump the \v Cerenkov photons are produced mainly
from scattered electrons (see section 7.2 for more details) and hence their 
fluctuations decrease with increasing primary energy.

It can be seen that the magnitude of fluctuations for proton primaries is
higher by a factor of 3--5 compared to those for $\gamma -$ ray primaries.
The difference seems to increase with increasing energy. At a $\gamma -$ ray
energy of 50 GeV this ratio is close to 3 while it is around 5 for 1 TeV
$\gamma -$ rays. The large rms fluctuations in the case of proton primaries
are mainly due to the increased relative fluctuations in the electron
number in the atmosphere. This is much more significant compared to the
Poissonian component. Further, the near constant relative fluctuations
show that the contribution from correlated emission from high energy
electrons is small (see section 7.1 for more details).

\begin{table} 
\caption{The average production height (km) of \v Cerenkov photons at a given
core distance range resulting from vertically incident $\gamma -$ rays of three
different energies at the top of the atmosphere. At near core distances the
\v Cerenkov photons are produced at relatively lower heights especially
at higher energies through direct emission of \v Cerenkov photons by the
surviving electrons.}
\begin{tabular}{llll} \hline
 {\bf Core distance (m) }  &  {\bf 100 GeV }  &  {\bf 500 GeV} & {\bf 1000 GeV} \\ \hline
     30-60        &  9.0       & 7.3 & 6.5 \\ 
    115-130        &  10.6     &  9.3   & 8.6  \\ 
    220-250        & 9.9       & 8.5  &  7.8 \\ \hline
\end{tabular}
\end{table}

\begin{table} 
\caption{The average angle of arrival (deg.) of \v Cerenkov photons at
different core distance ranges, resulting from vertically incident 
$\gamma -$ rays of
three different energies at the top of the atmosphere. At near core distances
the mean arrival angles increase with increasing primary energy because of
direct emission of \v Cerenkov photons by the
surviving electrons, which have a larger lateral spread.}
\begin{tabular}{llll} \hline
 {\bf Core distance (m) }  &  {\bf 100 GeV }  &  {\bf 500 GeV} & {\bf 1000 GeV}  \\ \hline
     30-60        &  0.38       & 0.47 & 0.53 \\ 
    115-130        & 0.73       &  0.81   & 0.87  \\ 
    220-250        & 1.46       & 1.73  & 1.91  \\ \hline
\end{tabular}
\end{table}

Figure 7 shows the ratio of the observed total RMS fluctuations (as against
non-statistical fluctuations in fig. 6) to the expected fluctuations if 
they were purely Poissonian, for $\gamma -$ rays and protons as a function of 
core distance. For $\gamma -$ rays, observed fluctuations
are larger than Poissonian at all energies upto the hump, approaching
Poissonian value beyond the hump, although they are always higher than
Poissonian in the core distance range considered here. In case of protons,
relative fluctuations are much larger than those for $\gamma -$ rays and 
decrease with increasing core distance.
In this case too the tendency to converge to Poissonian
fluctuations at large core distance is perhaps present probably at much larger core
distances. For $\gamma -$ ray primaries relative RMS fluctuations seem to be 
independent of energy of the primary beyond the hump, whereas in case of 
protons the fluctuations increase with increase in energy.
In summary, the density fluctuations are significantly non-Poissonian
and more prominent at near core distances.

The relative fluctuations shown in fig. 7 are not independent of the
mean density. It can be shown that

\begin{equation}
\sigma^T_{RMS}/\sigma^P = \sqrt{(1+\sigma^{NS}_{RMS}r)}
\end{equation}

where $\sigma^T_{RMS}$ and $\sigma^{NS}_{RMS}$ are the total and 
non-statistical RMS fluctuations and $\sigma^P$ is the Poissonian 
fluctuations given by the square root of the mean density. 
Fig. 7 shows only the relative fluctuations for two
types of primaries.

It has been suggested in the past
(Vishwanath, et al., 1993) that these non-statistical fluctuations could be
measured in an observation and used to improve the signal to noise
ratio. However, it was routinely assumed that for $\gamma -$ ray
primaries the fluctuations are Poissonian (Rao \& Sinha, 1988; Hillas and
Patterson, 1987). Such conclusions based on the assumption of Poissonian
fluctuations in the photon densities in the lateral distribution have to be
revised.

\section{Other statistical parameters}

\begin{figure}
\centerline{\psfig{file=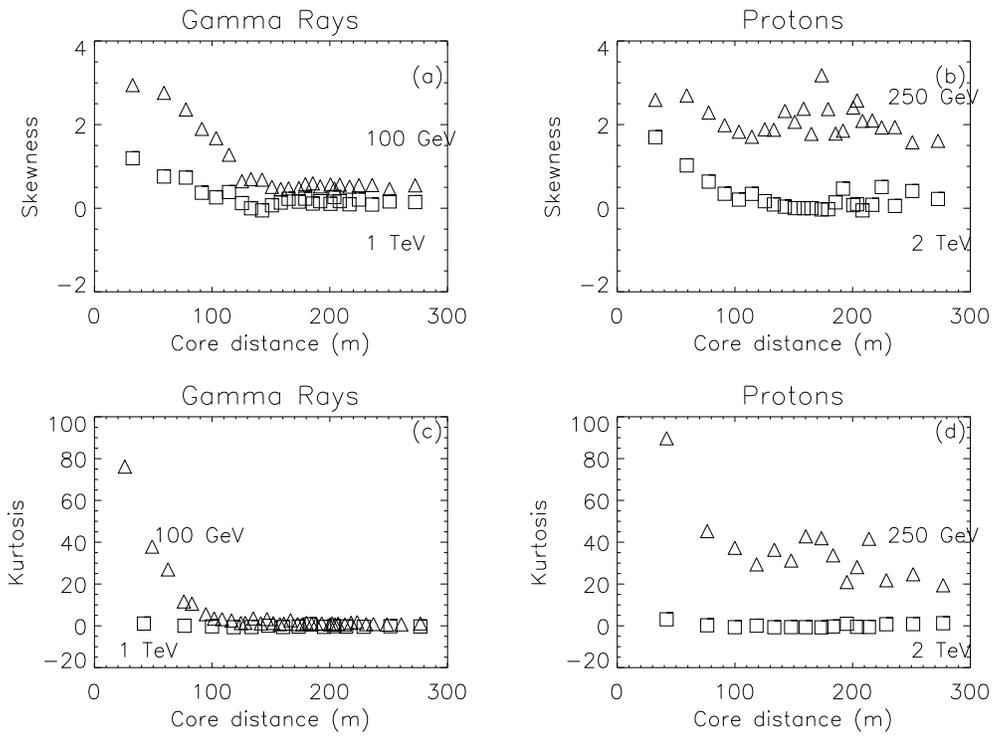,height=10cm}}
\caption{Variations of the third and fourth moments of the \v Cerenkov photon
density fluctuations as a function of core distance at two different energies  
for (a) \& (c) $\gamma -$ rays and (b) \& (d) protons.} 
\label{Figure 8:}
\end{figure}

Figure 8 shows the variation of the higher statistical moments like the 
skewness (a \& b) and kurtosis (c \& d) of the frequency distribution of
\v Cerenkov photon densities as a function of core distance for $\gamma -$ ray 
(a \& c) and proton primaries (b \& d). Each of the plots has two curves
for two different primary energies as indicated. For $\gamma -$ ray
primaries the density fluctuations show finite positive skewness before the
hump region showing that the distributions have a longer tail towards the
higher densities. The observed skewness is more than the statistical value of
$1 \over {(mean)^{1 \over 2}}$.
The \v Cerenkov photons
within the hump region are contributed to by two different sources: (i) 
electrons past the shower maximum emitting \v Cerenkov photons directly
before undergoing scattering,
which fill the region between core and the hump and (ii) \v Cerenkov photons
emitted by lower energy electrons after one or more Coulomb scatterings.
Larger skewness seen in case of lower energy $\gamma -$ ray primaries is
because of the relative fluctuations in the number of electrons surviving below
shower maximum. For lower energy $\gamma -$ rays the mean number of surviving
electrons is much 
smaller compared to higher energy $\gamma -$ rays and hence subject to
larger fluctuations resulting in larger skewness. At smaller mean values the
electron number distribution is asymmetric leading to a large skewness. 
After the hump the skewness becomes zero as the distribution becomes more 
symmetric. This could be understood since the photons beyond the hump
region are mainly from Coulomb scattering of low energy electrons. In
addition, this dependence of the skewness on the core distance gets diluted
at higher primary energies. For proton primaries, however, there is no marked
core distance dependence as the distribution has rather large positive
skewness uniformly at all core distances at lower primary energies, which
seem to decrease fast with increasing energy. In addition, isolated high 
energy electron tracks give rise to excessive photon densities (section 7.1)
leading to a large positive skewness.

Similarly the distributions at lower primary energies seem to show a
positive non-statistical kurtosis, indicating that the distributions are 
sharply peaked
compared to normal distribution and this effect vanishes at higher energies.

\section{Possible Origin of fluctuations}

There are several possible sources of \v Cerenkov photon density
fluctuations as seen at an observation level which is relatively far
from the shower maximum.

\subsection{Non-independent production processes}

 A single relativistic electron track can
emit several \v Cerenkov photons which are generically related and hence the
conventional statistics does not apply because of lack of independence among 
these photons. These \v Cerenkov photons
are strongly correlated and consequently give rise to large
non-Poissonian fluctuations at the observation level (Sinha, 1995).
Occasional local anomaly could be caused by high energy electron track
accentuating the above effect. This is demonstrated in fig. 9, which
shows the number of detected \v Cerenkov photons in (a) pre-hump, (b) hump
and (c) post-hump regions, at distances of 32 m, 123 m and 234 m from the
core, for 100 showers produced by $\gamma -$ ray primary with energy of
500 GeV. Also shown in the figure is average electron energy at 
atmospheric depth of 400 gm cm$^{-2}$, which is just below shower maximum
(at about 320 gm cm$^{-2}$).
Higher average energy of electrons seen in shower no. 38 and 49 is correlated
with significantly higher photon densities in pre-hump region. Hump and 
post-hump regions do not have such significant excess in photon densities. 
Thus showers with larger average energy for electrons 
produce distinctly higher density of \v Cerenkov photons only in pre-hump
region. This explains the larger fluctuations seen in fig. 5 for all 
energies of $\gamma -$ ray primaries in pre-hump region. Fig. 10 shows
the correlation between \v Cerenkov photon density and average electron 
energies at an atmospheric depth of 400 gm cm$^{-2}$ as a function
of core distance. Larger correlation in pre-hump region is evident.
By tracing the lateral extent of these large density fluctuations it has 
been found that the linear length scale varies from about 80 m to 120 m
with a mean value of about 100 m. This characteristic length scale does
not seem to be a sensitive function of $\gamma -$ ray energy.

\begin{figure}
\centerline{\psfig{file=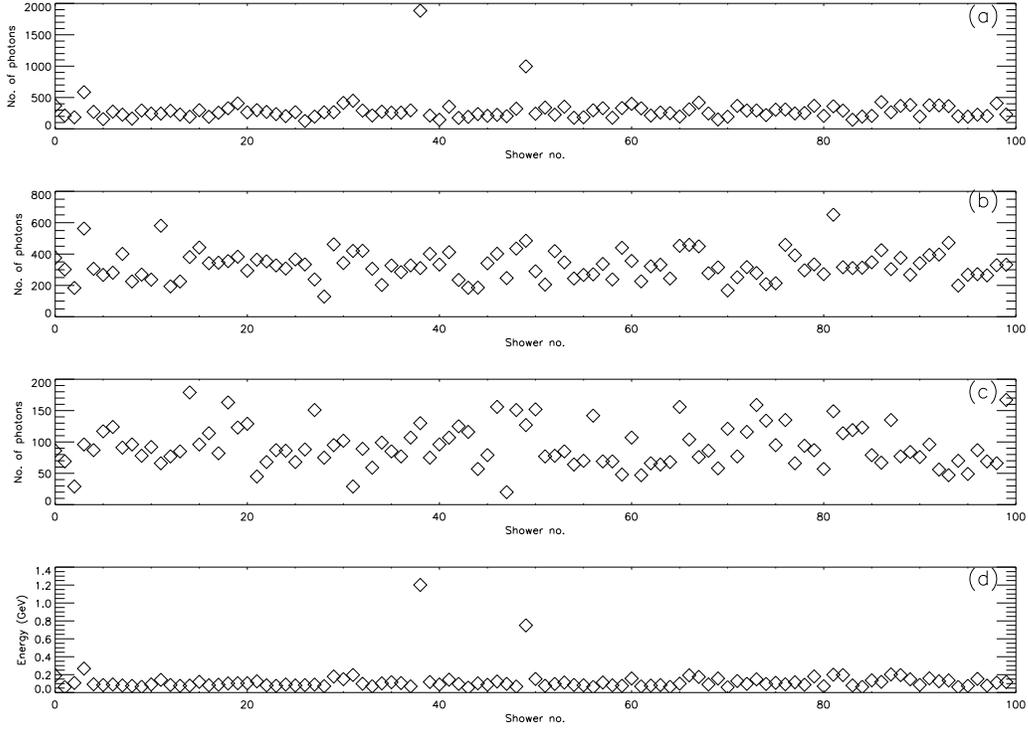,height=10cm}}
\caption{ This figure shows the total number of \v Cerenkov photons detected 
by a detector in
(a) pre-hump, (b) hump and (c) post-hump regions, at a distance of 32, 123
and 234m respectively from the core. Average energy of electrons at an
atmospheric depth of 400 gm cm$^{-2}$ is shown in (d). The primary $\gamma -$
ray energy is 500 GeV.}
\label{Figure 9:}
\end{figure}
  
\begin{figure}
\centerline{\psfig{file=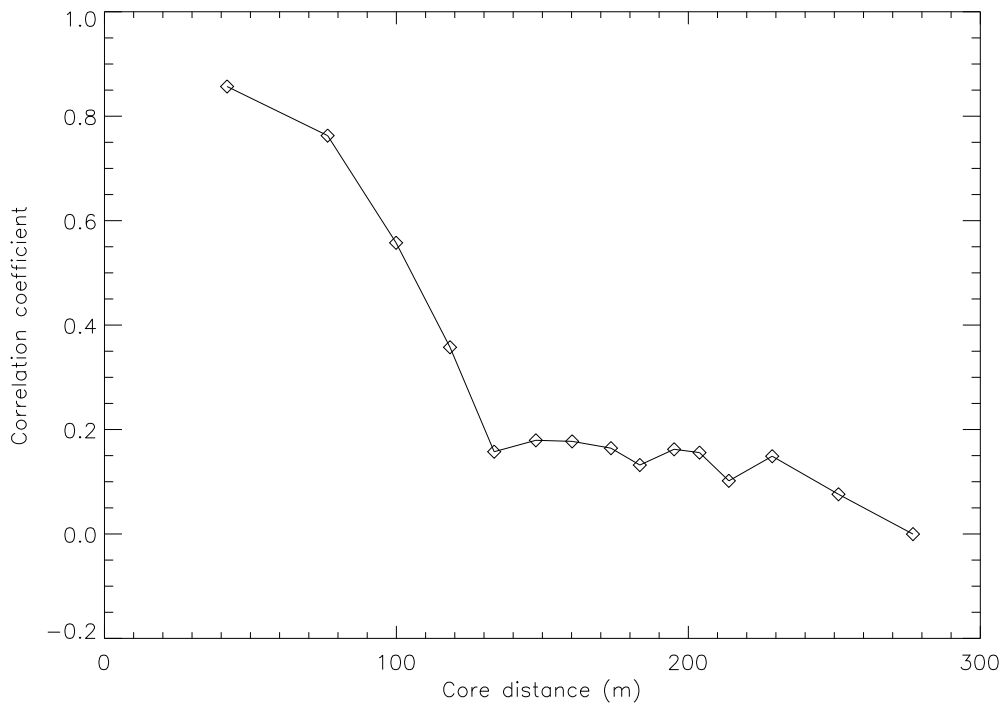,height=10cm}}
\caption{Correlation between number of detected \v Cerenkov photons per 
detector and average energy of electrons at atmospheric depth of 400 
gm cm$^{-2}$ as a function of core distance.
100 showers of 500 GeV $\gamma -$ rays were used. Points denote averages over
25 consecutive detectors when arranged in increasing order of core distance. 
}
\label{Figure 10:}
\end{figure}

\subsection{Coulomb scattering of electrons}

\begin{table}
\caption{The average angle of arrival of \v Cerenkov photons at a given core
distance resulting from vertically incident $\gamma -$ rays of energy 100 GeV
at the top of the atmosphere. This, when compared with the maximum core
distance of
direct \v Cerenkov photons, demonstrates the need for Coulomb scattering of low
energy electrons to account for the observed photon densities
(see text for details).}
\begin{tabular}{lllllll} \hline
 {\bf Core }  &  {\bf Aver. }  &  {\bf Atm. } &\bf Atm.& {\bf Atms }  &  {\bf \v
 Ceren.}  &  {\bf Max.} \\
 {\bf Dist. }  &  {\bf angle}  &  {\bf Height }  &  {\bf Height } & {\bf Depth }
  &  {\bf Angle}  &  {\bf core} \\
 & $(^{\circ })$   &  {\bf (derived.)}&  {\bf (sim.)} & {\bf $g$ $cm^{-2}$} &  {
\bf  $(^{\circ })$}  &  {\bf dist.} \\
 &   &  {(km) }&  { (km) } &  &   &  { (m)} \\ \hline
     30-60 m       &  0.38       & 6.94 &  8.96 & 421       &  0.94       & 114.
0 \\
    115-130 m       &  0.73       &  9.42   &  10.6 & 293       &  0.81       &
 132.3 \\
    220-250 m       &  1.46       & 9.3  &  9.88  & 299       &  0.81     &  131
.8 \\ \hline
\end{tabular}
\end{table}

\begin{table}
\caption{The average angle of arrival of \v Cerenkov photons at a given core
distance resulting from vertically incident protons of energy 250 GeV at the
top of the atmosphere. This, when  compared with the maximum core distance
 of direct
\v Cerenkov photons, demonstrates the need for Coulomb scattering of low energy
electrons to account for the observed photon densities
(see text for details).}
\begin{tabular}{lllllll} \hline
 {\bf Core }  &  {\bf Aver.}  &  {\bf Atm. } &\bf Atm.& {\bf Atm. }  &  {\bf \v
Ceren.}  &  {\bf Max.} \\
 {\bf Dist.}  &  {\bf angle}  &  {\bf Height }  &  {\bf Height } & {\bf Depth}
&  {\bf Angle}  &  {\bf core} \\
 &  $(^{\circ })$   &  {\bf (derived) }&  {\bf (sim.) } & {\bf $g$ $cm^{-2}$} &
 {\bf $(^{\circ })$}  & {\bf dist.} \\
 & &  { (km)}&  { (km)} &  &   & (m) \\ \hline
   30-60 m       &  0.60       & 4.43 &  6.5 & 593       &  1.08     & 83.7  \\
    115-130 m    &  0.96     &  7.15   &  6.96 & 409     &  0.93     & 116 \\
  220-250 m      &  1.46     & 9.3  &  7.58  & 299       &  0.81     &  131 \\ 
\hline
\end{tabular}
\end{table}

The Coulomb scattering of low energy electrons can lead to 
density fluctuations as the \v Cerenkov photons that are emitted by an
electron could be diverted to a different location on the ground due to
scattering of the parent electron in air. However in general Coulomb
scattering tends to smear the fluctuations caused by processes mentioned
in section 7.1. Table 4 demonstrates the effect of
scattering on the observed lateral distribution of \v
Cerenkov photons in the case of a $\gamma -$ ray primary of 100 GeV incident
vertically at the top of the atmosphere. The second column lists the average
observed arrival angle of \v Cerenkov photons at a core distance range given in
column 1. The core distances were suitably chosen to represent 
pre-hump, hump and post-hump regions respectively. 
Column 3 lists the atmospheric height where the \v Cerenkov photons 
arriving at the angle shown in column 2 could have originated. Column 4 
shows the average atmospheric height, estimated from the simulation results, 
which contributes \v Cerenkov photons at core distance range shown in
column 1. Column 5 shows the atmospheric depth at the height shown in
column 3 and column 6 shows the \v Cerenkov angle at that height. Column 7 
is maximum core distance at which \v Cerenkov photons reach the observation 
level.  
Only at the hump region the core distances of the
observed photons (column 1) and those of the direct hit \v Cerenkov photons 
(column 7)
agree well indicating that these photons are emitted by electrons without
undergoing significant Coulomb scattering. While the directly emitted 
\v Cerenkov photons would not have reached the 
observed core distances in pre- and post-hump regions if the emitting
electrons had not undergone Coulomb scattering significantly.
The difference between the derived and observed effective production heights 
of \v Cerenkov photons in pre-hump and post-hump regions (columns
3 and 4) again demonstrates the dominance of Coulomb scattering of electrons.

Table 5 gives a similar information for proton primaries of energy 250 GeV.
Here also the photons at a core distance of around 120 m have their \v 
Cerenkov angles and the arrival angles equal. At shorter core distances
the average arrival angles are larger than those for $\gamma -$ ray primary 
because of the larger lateral spread of the electrons in the case of proton 
primaries. However
the need for significant Coulomb scattering of electrons is borne out by
the difference in the maximum reach of the \v Cerenkov photons and the
observed core distance.

\subsection{Fluctuations in the height of first interaction}

The height of first interaction of a primary proton or a $\gamma -$
ray in the atmosphere that initiates the cascade fluctuates within one
interaction length. The extent of fluctuation is decided by the radiation 
length in air ($37.15 g cm^{-2}$)in the case of $\gamma -$ ray primaries and
the interaction mean free path ($70 g cm^{-2}$) in the case of protons. Fluctuations in the
point of first interaction in turn lead to the fluctuations in the height
of the shower maximum. A shower with a maximum at a point lower down in the
atmosphere effectively comes closer to the observation level and hence
results in higher number of \v Cerenkov photons since the \v Cerenkov yield is
an increasing function of refractive index. Thus fluctuations in the height
of the shower maximum give rise to density fluctuations at the observation
level. Both the height of first interaction as well as the height of shower
maximum for a primary $\gamma -$ ray of given energy is correlated
with shower size (see table 6). 
However, such a correlation is absent in the case of proton
primaries. Table 7 lists the magnitude of fluctuation of these parameters
for a 100 GeV $\gamma -$ ray incident vertically at the top of the
atmosphere.

\begin{table} 
\caption{Correlation coefficients of the shower size with the height of 
shower maximum and height of first interaction point from a sample of 300 
showers each of $\gamma -$ ray and proton primaries.}
\begin{tabular}{lll} \hline
     {\bf
Primary species}  &  {\bf Corr. Coeff. of}  &  {\bf Corr. Coeff. of} \\ 
  
&  {\bf shower size with }  &  {\bf shower size with} \\
 &  {\bf shower}  &  {\bf height (km) of first} \\ 
 &  {\bf maximum ($g$ $cm^{-2} $)}  &  {\bf interaction} \\ \hline
       $\gamma -$ rays (100 GeV)   &  0.79       &  -0.7  \\  
       $\gamma -$ rays (1000 GeV)      & 0.83       &  -0.71  \\  
     Protons  (250 GeV)     &  0.16       & -0.11  \\ 
 Protons  (2000 GeV)     &  0.23       & -0.26  \\
\hline
\end{tabular}
\end{table}

\begin{table} 
\caption{Magnitude of fluctuations in the height of first point of interaction, 
shower maximum and the number of \v Cerenkov photons detected at
observation level for
a 100 GeV $\gamma -$ ray incident vertically at the top of the atmosphere.}
\begin{tabular}{llll} \hline
     {\bf Parameter}  &  {\bf Mean }  &  {\bf RMS } & {\bf Relative RMS}\\
         & {\bf value } & {\bf fluctuations} & {\bf fluctuations} \\
 \hline
     Height of first interaction (m)       &  25582.2    & 7962.3  & 31.1\% \\ 
    Height of shower maximum   &  260.6   &  65.70  & 25.3\%\\ 
 ($g$ $cm^{-2} $) & & & \\
   No. of \v Cerenkov photons detected    &  11480   &  3050  & 26.6\%  \\
  at the observation level        &       &    \\ \hline
\end{tabular}
\end{table}

The \v Cerenkov photon density fluctuations at the observation level is a
result of multiple components. While it is not possible to quantify the
contributions from all the processes separately, one can estimate the
contribution from the fluctuations in the height of shower maximum.
A sample of 500 $\gamma -$ray showers of energy 100 GeV have been simulated
and  a distribution (bin width = 1 km equivalent to $\sim$ 4 gm cm$^{-2}$ at
an altitude of $\sim$ 25 kms) of 
these showers is generated based on their height of
first interaction in the atmosphere.
The mean RMS value of the fluctuations of
the shower size in the bins around the peak of 
is distribution is computed to be 3\%, compared to 5.2\% for the totality
of the showers. This value is presumably free from the contribution
to fluctuations from the fluctuations in the height of first interaction. 
This value does
not vary significantly if we reduce the bin width by a factor of 2. Hence the
fluctuations in the height of first interaction contributes significantly to
the observed fluctuations in the shower size for $\gamma -$ ray primaries.

A similar estimate for proton primaries shows that the contribution from the
fluctuations in the height of first interaction is negligible showing that 
the contributions from other processes dominate.
In the case of proton primaries, it was possible to freeze the height of
first interaction at the mean value for 250 GeV proton primaries. 
The value of the RMS fluctuations in the shower
size after fixing the height of first interaction is 29.2\%, compared to
30\% obtained without doing so, confirming the above conclusion.

\subsection{Electron number fluctuations}

In an attempt to further understand the relative contributions to the observed
density fluctuations at the observation level, we studied the fluctuations
in the production height of \v Cerenkov photons for $\gamma -$ ray and
proton primaries. Fig. 11(a) shows the production height distribution of \v
Cerenkov photons generated by 100 GeV $\gamma -$ rays, averaged over 50
showers. Error bars correspond to RMS deviations and are clearly
non-statistical. Electron growth curve is also shown in the fig 11(c). The
photon and electron growth curves are very similar as expected. Both 
the number of \v
Cerenkov photons and electrons peak at a height of around 10 km, which is the
shower maximum. Fluctuations in the number of \v Cerenkov photons at a given
height in the atmosphere can be attributed to those in the number of
electrons, which in turn owe their origin to the production kinematics. 
It may be noticed from plots in 11(b) \& (d) where relative rms
deviations are plotted as a function of height, that the fluctuations are
relatively larger after the shower maximum. 

\begin{figure}
\centerline{\psfig{file=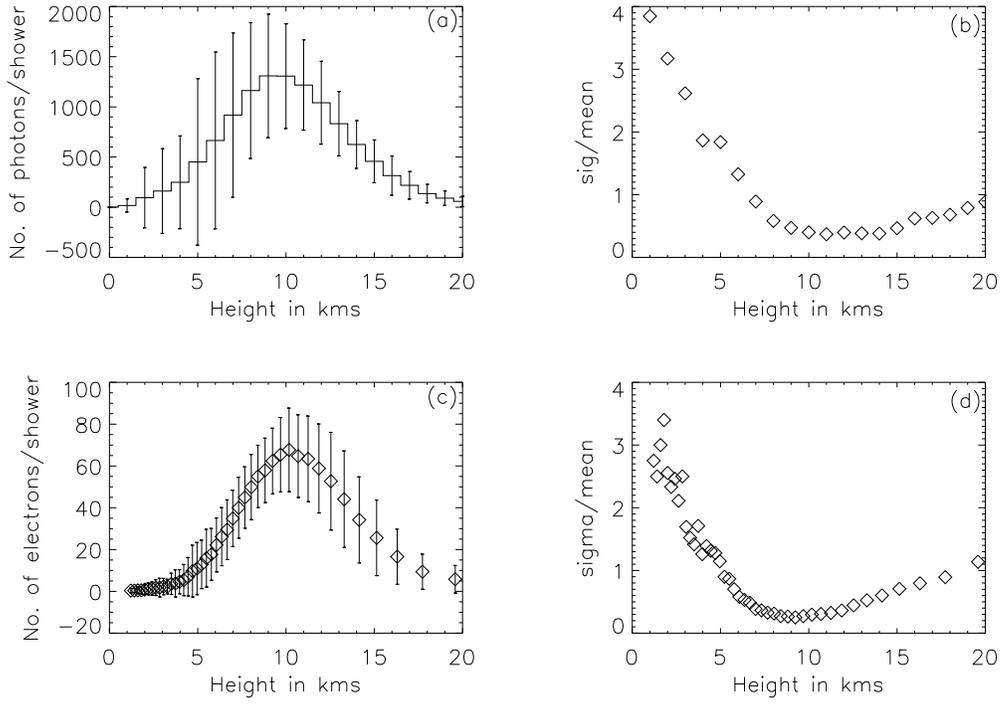,height=10cm}}
\caption{Production height distribution of (a) \v Cerenkov photons and (c)
electrons generated by 100 GeV $\gamma -$ ray primaries. The error bars
indicate the rms deviations. The plots on the right show the relative errors
measured as a ratio of the average number (over 50 showers) for (b) photons and
(d) electrons. The number on the y-axis of (a) and (c) could be multiplied by
2 to include positrons as well.}
\label{Figure 11:}
\end{figure}

\begin{figure}
\centerline{\psfig{file=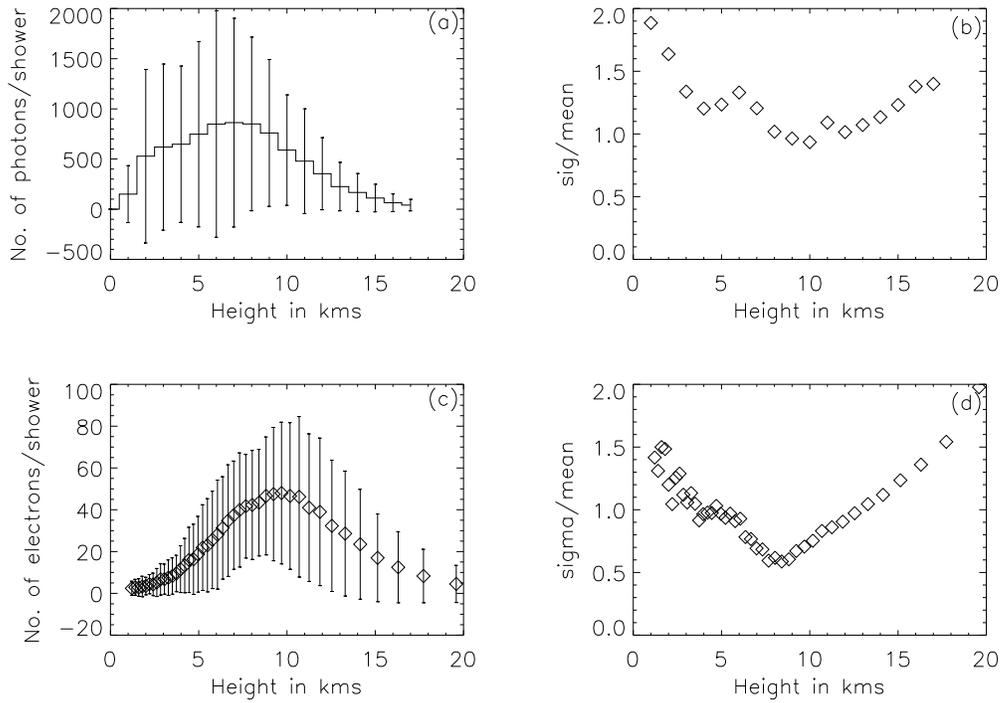,height=10cm}}
\caption{Production height distribution of (a) \v Cerenkov photons and (c) 
electrons generated by 250 GeV proton primaries. The error bars indicate 
the rms deviations. The plots on the right show the relative errors measured 
as a ratio of the average number (over 50 showers) for (b) photons and
(d) electrons. The number on the y-axis of (a) and (c) could be multiplied by
2 to include positrons as well.} 
\label{Figure 12:}
\end{figure}

Similar distributions averaged over 50 showers of protons of energy 250 GeV
are shown in fig. 12. Here, it may be noticed that the magnitude of
fluctuations is much larger compared to those of $\gamma -$ rays. 
This is because in the case of proton primaries, the source of electrons is
pair production by $\gamma -$ rays which are the decay products of $\pi ^0$
mesons produced in hadronic interactions. While charged pions could decay
to muons which in turn decay to electrons. The fluctuations in the
multiplicities of the pion secondaries and their energy spectra combined 
with the fluctuations due to larger hadronic interaction mean free path 
(which is nearly twice the radiation length in air) of protons lead to
larger electron number fluctuations in the case of proton primaries. 
The shower development also sustains over a longer distance in the
atmosphere for the same reason. The
FWHM of the shower development curve for $\gamma -$ ray and proton primaries
are 8 and 10 km respectively. This is due to the finite transverse momentum
of $\pi ^0$ secondaries which is about 0.3 GeV/c in the case of proton
primaries.

\subsection{Electron energy spectra at various atmospheric depths}

\begin{figure}
\centerline{\psfig{file=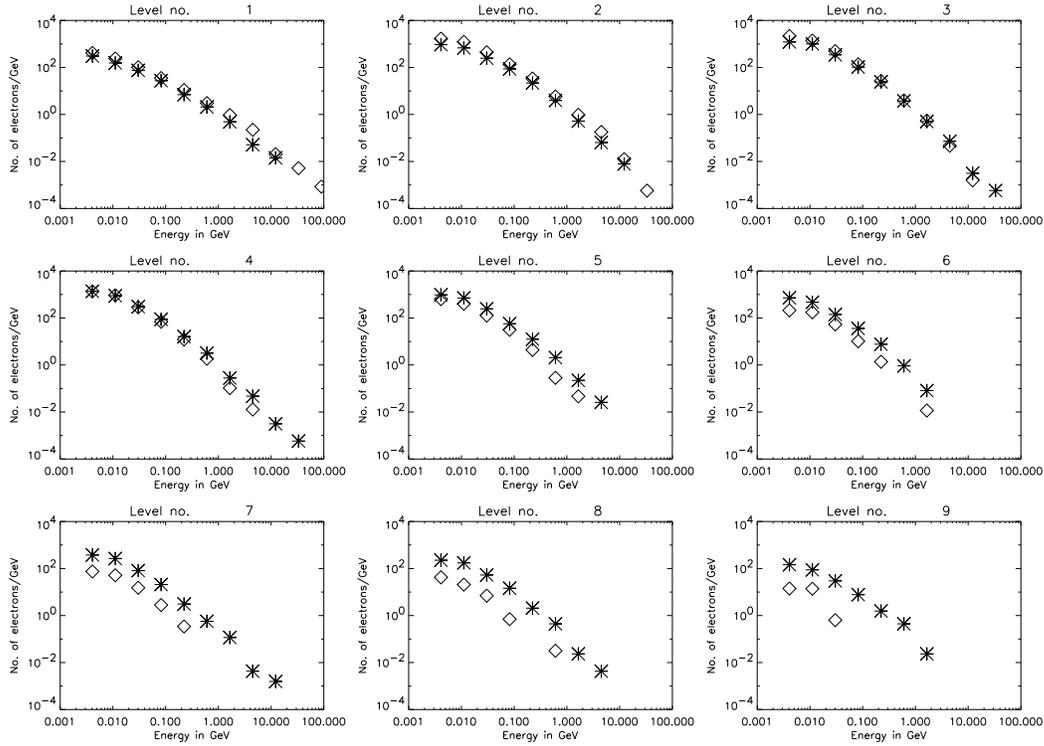,height=10cm}}
\caption{Averaged electron energy spectra (diff) at 9 different atmospheric depths 100-900 $g$ $cm^{-2}$
during the shower development for $\gamma -$ ray(diamonds) and proton(stars) primaries of energy 100 \& 250
GeV respectively.} 
\label{Figure 13:}
\end{figure}

Another important contribution to fluctuations comes from the fluctuations in
the electron energy and their spectra during the shower development. Figure 13 
shows the average energy spectra of electrons (averaged over 50 showers) at 9
different depths (100 - 900 $g$ $cm^{-2}$) in the atmosphere, both for proton
(250 GeV, stars) and $\gamma -$ ray primaries (100 GeV, diamonds), incident
vertically at the top of the atmosphere. Figure 14(a) shows the variation of
the average energy of the electrons (calculated as the arithmetic mean of all 
the electron energies), as a function of atmospheric depth. While the energy
spectra seem to be very similar in shape for $\gamma -$ ray and proton
primaries, the average energy after the shower maximum is relatively higher 
for proton primaries since their height of shower maximum is past that for $%
\gamma -$ ray primaries. For the same reason the number of surviving
electrons also is relatively larger in the case of proton primaries. The
relative fluctuations in the average energy shown in fig. 14(b) indicate the
RMS fluctuations which propagate down to those in the number of \v Cerenkov 
photons at the observing level. It may be noted that the relative RMS
fluctuations in the average electron energies at various atmospheric depths
are very similar in the case of $\gamma -$ ray and proton primaries. Hence
the increased fluctuations in the \v Cerenkov photon densities for proton
primaries is mainly the result of increased fluctuations in the electron 
number during the shower development.

\begin{figure}
\centerline{\psfig{file=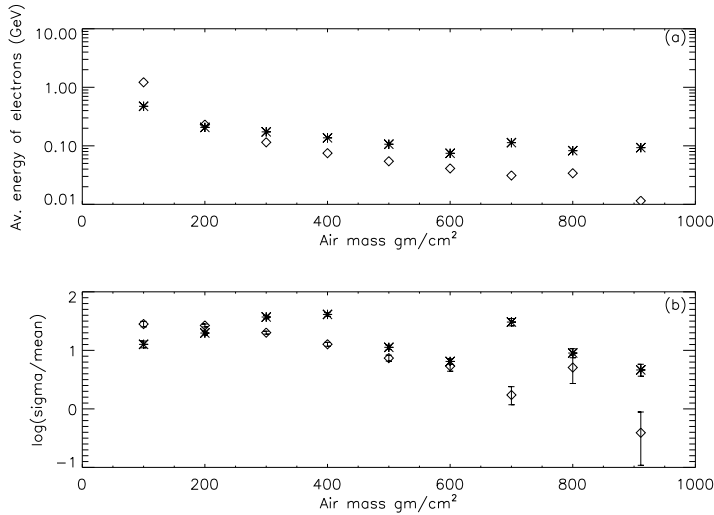,height=7cm}}
\caption{(a) Average electron energy  at 9 different atmospheric depths 
100-900 $g$ $cm^{-2}$ during the shower development for $\gamma -$ ray 
(diamond) and proton (star) primaries of energy 100 \& 250 GeV respectively.
(b) Ratio of RMS deviation to average energy for $\gamma -$ ray and proton
primaries.} 
\label{Figure 14:}
\end{figure}

\section{Comparison with other simulations}

\subsection{Average lateral distributions}

\begin{figure}
\centerline{\psfig{file=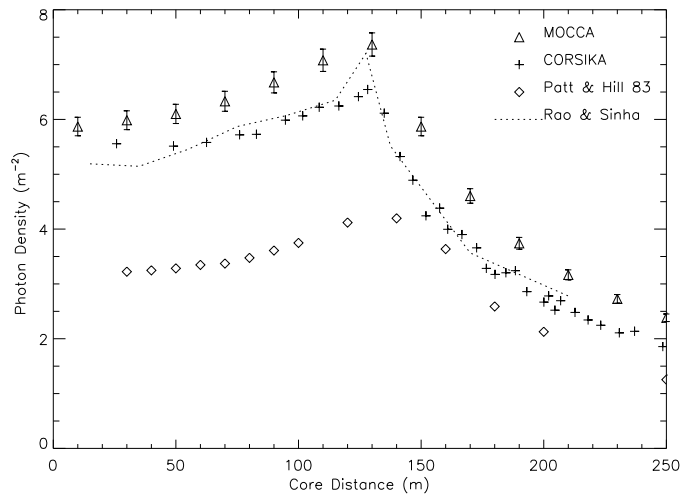,height=7cm}}
\caption{Average lateral distributions of \v Cerenkov photon densities
resulting from vertically incident $\gamma -$ rays of energy 100 GeV. 
The average of 200 
showers are shown as computed by CORSIKA, MOCCA, Rao \& Sinha simulation  
package and Patterson \& Hillas calculations.}
\label{Figure 15:}
\end{figure}

Here we compare the average (over 200 showers) lateral distributions of \v
Cerenkov photons for $\gamma -$ ray primaries of energy 100 GeV obtained
using CORSIKA, with (a) that obtained by Ong et al. (1995) using a
package called MOCCA (Hillas, 1985) (b) a package developed by Rao \& Sinha
(1988) and (c) the earlier calculations by Patterson \& Hillas (1983). The
average lateral distributions of Patterson \& Hillas were originally
made for sea level. The expected density for Pachmarhi altitude (using the
estimates of Rao \& Sinha (1988)), is around 30\% higher than that for
sea-level. However, the core distance of the hump also reduces at higher
altitudes and there is no easy way to correct for this effect. Results
are shown in the fig. 15. We have corrected the lateral distribution
obtained with CORSIKA for atmospheric attenuation of \v Cerenkov photons
using the same attenuation factor as used by Rao \& Sinha (Acharya, 1997). All
the three lateral distributions (a), (b) and (c) agree within errors while
(d) agrees with the rest only beyond the hump region. It may be mentioned
that the estimates by Rao \& Sinha do not take into account the effects
of geomagnetic field. Slightly higher photon density obtained by Ong et al.
may be due to the differences in \v Cerenkov bandwidth and altitude used by
them. The calculations of Patterson \& Hillas (1983) are underestimated by
about 22\% compared to CORSIKA, even after accounting for the differences in
the observation levels. The photon densities as well as the hump agree
reasonably well in cases a, b and c, showing that all the three simulation 
packages include correct treatment of Coulomb scattering of high energy 
electrons and their relative track length integrals.

\subsection{Average shower size}

As seen in section 4 the shower size (measured in terms of the total number of 
\v Cerenkov photons produced) for 100 GeV $\gamma -$ ray
primaries obtained using CORSIKA is about 4.3 $\times$ 10$^6$
photons and the ratio of RMS deviation to the average number of  
\v Cerenkov photons is 5\%. For the same $\gamma -$ ray energy Ong et al. 
(1995) have estimated the total number of \v Cerenkov photons within
150 m from the shower axis to be around 4.78 $\times$ 10$^5$ and the ratio
of RMS deviation to the number of \v Cerenkov photons to be 34\%. The 
difference between the numbers are mainly due to the difference in collection
area, which is a circle of radius 150 m centered at the core in case of Ong 
et al., whereas the numbers given by us correspond to the entire pool. Using 
detected number of \v Cerenkov photons at various core distances and using 
a quadratic fit to the lateral distribution, we estimate the number of 
\v Cerenkov photons within 150 m from shower axis to be around 6.27 $\times$ 
10$^5$. Considering the atmospheric attenuation it reduces to 3.49
$\times$ 10$^5$ photons, which is consistent with the number given by Ong et 
al. Also using the number of detected \v Cerenkov photons we have  
estimated the ratio of RMS deviation to the number of \v Cerenkov photons 
within 150 m of shower axis to be about 28\%. Considering an unequal radial 
distribution of the detector array used in our calculations, this number is 
also consistent with that obtained by Ong et al.

\subsection{Effect of geomagnetic field}

\begin{figure}
\centerline{\psfig{file=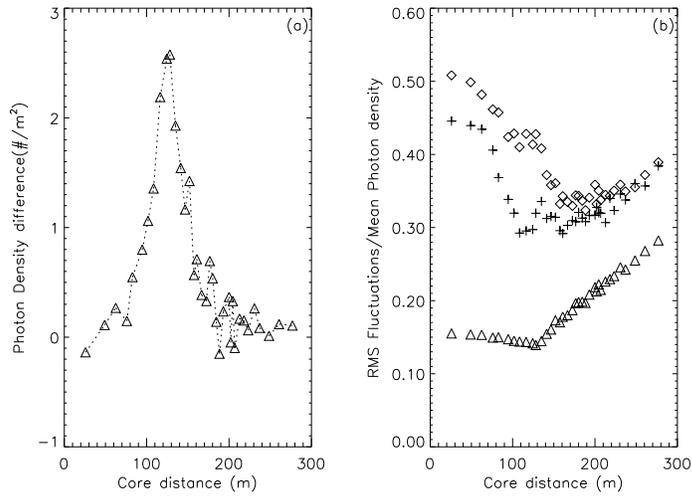,height=7cm}}
\caption{ (a) The effect of the geomagnetic field on the 
average lateral distribution of \v Cerenkov photons. The difference between 
the \v Cerenkov photon densities without and with geomagnetic field is 
plotted as a function of core distance. The increased prominence of the hump 
in the absence of the field is clearly seen. The primary $\gamma -$ ray 
energy is 100 GeV. See text for details.
(b) The change in the relative fluctuations of \v Cerenkov
photon densities as a function of core distance due to the presence of
geomagnetic field. Relative fluctuations with geomagnetic field are indicated
by diamonds, without field by + and Poissonian fluctuations by triangle.} 
\label{Figure 16:}
\end{figure}

It has been found that the effect of wavelength dependent absorption of \v
Cerenkov photons in the atmosphere is independent of core distance and hence
it does not affect the prominence of the hump. The difference between the
average lateral distributions derived from CORSIKA taking into account the 
local geomagnetic field and without taking into account, is shown in
figure 16(a). It shows that the deflection of electrons in the Earth's
magnetic field would dilute the prominence of the hump by broadening it as
expected (Porter, 1973). The electron deflection in the magnetic field is
equivalent to increased Coulomb scattering and hence is expected to produce
more fluctuations in the observed photon densities. Figure 16(b) shows the
relative rms fluctuations with and without taking into account the presence
of the geomagnetic field confirming the above conclusion. Also shown in the
plot are relative RMS fluctuations if they were purely Poissonian in origin.

\subsection{Density fluctuations}

There are very few results available in the literature on the study of
fluctuations of \v Cerenkov photons. It has been known qualitatively that
the shower to shower variations are relatively higher in the case of showers
initiated by hadrons compared to those generated by $\gamma -$ rays. This is
mainly because the electromagnetic cascades, which contribute to \v Cerenkov
photon density in the case of hadronic primaries, are the result of a
superposition of several $\gamma -$ ray cascades initiated by the decay products
of $\pi ^0$ mesons whose numbers are subjected to fluctuations. However, no
serious attempt was made to compare these fluctuations quantitatively and
qualitatively in the past.

Sinha (1995) studied these fluctuations only for $\gamma -$ ray primaries of
energies 100, 500 and 2000 GeV. There is a qualitative agreement between
these calculations and the present work in the sense that the degree of
fluctuations decrease from the core going through a minimum at around the
hump region and then showing a slight increase at large core distances. As
the photons at around the hump are contributed mainly by higher energy
electrons ($E_e>1\ GeV$) they will undergo relatively less scattering
and hence are subjected to
minimum fluctuations. At small core distances the asymmetric distributions
of densities lead to larger fluctuations, while at large core
distances the residual fluctuations seen are due to Coulomb scattering. 
Since the estimates by Sinha are for
sea level the degree of fluctuations cannot be compared quantitatively. Our
estimates of relative fluctuations as a function of core distance shows no
agreement with the Poissonian fluctuations at any core distance while Sinha
(1995) observes that beyond the hump region the observed fluctuations are
consistent with Poissonian. Since the Coulomb scattering of low energy
electrons is responsible for \v Cerenkov photons reaching beyond hump region
one would expect to see the residual contribution to fluctuations from
Coulomb scattering as shown in the present work. 

\section{Conclusions}

A systematic study of the fluctuations in the \v Cerenkov photons generated 
by $\gamma -$ ray and proton primaries in the earth's atmosphere and detected 
at the observation level has been carried out. Such a quantitative estimate 
of the degree of fluctuations for the two types of primaries and the 
dependence on the core distance as well as primary energy has been done for 
the first time here. This type of study is important in planning observations 
of VHE $\gamma -$ ray sources based on the measurement of lateral 
distributions of \v Cerenkov photons, since these experiments are based on 
improving signal to noise ratio by rejecting the abundant charged particle 
background. The large non-statistical fluctuations reported here might reduce 
the efficiency of rejection.

We have studied the density fluctuations as a function of core distance
for various energies of $\gamma -$ ray and proton primaries in
50--1000 GeV range. Fluctuations are highly non-statistical and
decrease with increasing primary energy in both the cases. Proton
primaries show larger fluctuations compared to $\gamma -$ ray primaries
of corresponding energy. In case of $\gamma -$ ray primaries fluctuations
are minimum at the hump region and approach Poissonian beyond the hump
region. Proton showers in general show larger fluctuations than
$\gamma -$ ray primaries even in the case of shower size measured in terms of 
the total number of \v Cerenkov photons generated.

We have investigated various known sources of fluctuations and tried
to evaluate their relative contributions. Effect of geomagnetic field
is to deflect electrons and thereby increase the fluctuations. Average
electron energies and their spectra at different atmospheric levels during the
shower development are found to be similar in case of proton and $\gamma -$ 
ray primaries of equivalent \v Cerenkov yields. Whereas number
of electrons at various depths of shower development is found to vary
much more for protons compared to $\gamma -$ rays. We have also
studied the effect of variation in the first point of interaction on the 
shower size fluctuations. It is found to be significant only in the case of
$\gamma -$ ray primaries. 
Contribution to the fluctuations also comes from Coulomb 
scattering of low energy electrons past the shower maximum and the intrinsic 
correlation between the photons emitted by a single electron.

\section{Acknowledgments}

We would like thank Profs. K. Sivaprasad, B. S. Acharya, M. V. S. Rao and P.
R. Vishwanath for extensive and illuminating discussions during present work.
We would like to thank the anonymous referee whose interesting comments and
suggestions helped us improve the quality of this paper. 

\section{References}

\begin{description}
\item  Acharya, B. S., Private Communication, (1997)

\item  Bhat, P. N., {\it et al}., Proc. 25th ICRC, Durban, {\bf 5}, 105,
(1997)

\item  Browning, R. and Turver, K. E., Nuovo Cimento, 38A, 223 (1977)

\item  Chadwick, P. M. {\it et al., }Proc. 25th ICRC, Durban, {\bf 3}, 189,
(1997)

\item Fegan D., High Energy Astrophysics - Models and Observations from
MeV to EeV, Ed. J. M. Mathews, p107, World Scientific, (1994)
 
\item Hillas, A. M., Proc. 19th ICRC, La Jolla, {\bf 3}, 445, (1987)

\item  Hillas, A. M. and Patterson, J. R., Very High Energy $\gamma -$ Ray
Astronomy, Ed. K. E. Turver (Dordrecht: Reidel), 234, (1987)

\item  Knapp, J. and Heck, D., {\it EAS Simulation with CORSIKA, V4.502: A
User's Manual}, (1995)

\item  Krys, E. and Wasilewski, A., {\it Towards a Major Atmospheric \
Cerenkov Detector-II, }Ed: R. C. Lamb, Calgary, 199, (1993)

\item  Nelson, W. R., {\it The EGS4 Code System}, SLAC Report 265 (1985)

\item  Ong, R. A. et al., {\it Towards a Major Atmospheric \ Cerenkov
Detector-IV, }ed. M. Cresti, pp. 261, (1995)

\item  Patterson, J. R. and Hillas, A. M., J. Phys. G: Nucl. Phys., 9, 1433,
(1983)

\item  Porter, N. A., Lett. Nuovo Cimento, 8, 481, (1973)

\item  Punch, M., {\it et al}., Nature, 358, 477 (1992)

\item  Quinn, J. et al., Proc. 25th ICRC, Durban, {\bf 3}, 249, (1997)

\item  Rao, M. V. S. and Sinha, S., Phys. G: Nucl. Phys.,14, 811, (1988)

\item  Sinha, S., J. Phys. G: Nucl. Phys., 21, 473, (1995)

\item  Vacanti , G., et al., Ap J, 377, 467, (1991)

\item  Vishwanath, P. R. et al., {\it Towards a Major Atmospheric \ Cerenkov
Detector-II, }Ed: R. C. Lamb, Calgary, 115, (1993)

\item Weekes T. C. Physics Reports, {\bf 160}, 1, (1988)

\item  Zatsepin, V. I. and Chudakov, A. E., Sov. Phys.-JETP, 15, 1126, (1962)
\end{description}

\end{document}